\documentclass[12pt,fleqn]{article}
\usepackage{ifthen}                   % needed for Feyman-Slash macro
\usepackage{exscale}                  % correct scaling of math-symbols
\usepackage[intlimits]{amsmath}       % improved mathematical typesetting
\usepackage{amsfonts}
\usepackage{amssymb,amscd}
\usepackage[dvips]{epsfig}                   
\usepackage{array}
\usepackage{wrapfig}

\begin{document}

\setlength{\baselineskip}{20pt}

\psfull

{\small

\rightline{UNITUE--THEP--07/2002}
\rightline{http://xxx.lanl.gov/abs/hep-ph/0202202}}

\vspace{1cm}

\begin{center}
\begin{large}
{\bf Infrared Exponents and Running Coupling\\ 
of SU(N) Yang--Mills Theories}\\
\end{large}
\vspace{1cm}
{\bf 
C.\ S.\ Fischer$^{1 \dagger}$ and R.\ Alkofer$^2$
\\}
\vspace{0.2cm}
{Institute for Theoretical Physics, T\"ubingen University \\
Auf der Morgenstelle 14, D-72076 T\"ubingen, Germany
}
\end{center}
\vspace{1cm}
\begin{abstract}
\normalsize
\noindent
We present approximate solutions for the gluon and ghost propagators as well as
the running coupling in Landau gauge Yang--Mills theories. We solve the 
corresponding Dyson--Schwinger equations in flat Euclidean space-time without
any angular approximation. This supplements recently obtained results employing
a  four-torus, {\it i.e.}~a compact space-time manifold, as infrared regulator.
We confirm previous findings deduced from an extrapolation with tori of
different volumes: the gluon propagator is weakly vanishing in the infrared and
the ghost propagator is highly singular. For non-vanishing  momenta our
propagators are in remarkable agreement with recent lattice calculations.
\end{abstract}
\vspace{2.5cm}
\vfill

\noindent
\rule{5cm}{.15mm}

\noindent
$^\dagger$Supported by the European Graduate School Basel-Tuebingen. 
\\
$^1$E-Mail:chfi@axion01.tphys.physik.uni-tuebingen.de\\
$^2$E-Mail:reinhard.alkofer@uni-tuebingen.de

\newpage

\pagenumbering{arabic}

The non-perturbative structure of the gluon propagator is of fundamental
interest in Quantum Chromo Dynamics, and its knowledge would provide an
important input for many calculations in hadron physics \cite{Alkofer:2001wg}.
As infrared singularities might be encountered, a non-perturbative continuum
method seems mandatory. In this context, the Landau gauge Dyson--Schwinger
equations have been solved analytically in the infrared, but had to be
truncated in order to get a closed system of
equations \cite{vonSmekal:1997is,Atkinson:1998tu,Atkinson:1998cz,
Zwanziger:2001kw,Lerche:2001}.
On the other hand, lattice calculations  \cite{Bonnet:2000kw,Langfeld:2001cz}
include all non-perturbative physics of Yang-Mills  theories but are limited for
small momenta by the finite lattice volume. Despite these  shortcomings both
approaches agree surprisingly well in their general statements: there is clear
evidence for an infrared finite or even vanishing gluon propagator and a 
diverging ghost propagator. This is in accordance with the Kugo-Ojima
confinement criterion, which in Landau gauge includes the statement that the
ghost propagator should  be more singular than a simple pole
\cite{Kugo:1995km}.

One of the obstacles encountered in extracting definite values of infrared
exponents in Dyson--Schwinger studies is the angular integrals inherent to
these equations. Therefore approximated treatments of the angular integrals
have been applied in numerical works  so far
\cite{vonSmekal:1997is,Atkinson:1998tu}. In general the angular
approximations proved to be good for high momenta but less trustworthy  in the
infrared. Recent studies \cite{Atkinson:1998cz,Zwanziger:2001kw,Lerche:2001} 
therefore concentrated on the infrared analysis, where exact results have been
gained for the limit of vanishing momentum.

In a previous paper \cite{Fischer:2002} a truncation scheme has been presented,
which on the  one hand provides the correct anomalous dimensions of the ghost
and gluon dressing functions, $Z(k^2)$ and $G(k^2)$, in the ultraviolet region
of momentum. On the other hand, it reproduces the infrared exponents found in
\cite{Zwanziger:2001kw,Lerche:2001} which are close to the ones extacted from
lattice calculations \cite{Bonnet:2000kw,Langfeld:2001cz}.  The coupled set of
equations for the ghost and gluon dressing functions on  a four-torus have been
solved recently \cite{Fischer:2002}. This turned out to be an effective tool to
overcome the angular approximation for non-vanishing momenta. The numerical
solutions proved to be compatible with only one out of  two solutions of the
infrared analysis of refs.~\cite{Zwanziger:2001kw,Lerche:2001}, thus suggesting
that not every analytical solution for zero momentum connects to numerical
solutions for non-vanishing momenta. However, employing the torus as an
infrared regulator shares with lattice calculations the problem of working on a
finite  volume. Therfore further investigations are highly desirable. 

In this letter we present the solution of the coupled set of equations
described in ref.~\cite{Fischer:2002} in flat Euclidean space-time. The
results confirm our  previous findings on the torus. We obtain a
numerical solution for the gluon and the ghost propagator with infrared
behavior $D_{gl}(k^2) \sim (k^2)^{2 \kappa -1}, \kappa \approx 0.595$ and
$D_{gh}(k^2) \sim (k^2)^{-\kappa -1}$. Correspondingly we find an infrared
fixed point of the running coupling, $\alpha(0) \approx 2.972$ for the gauge 
group SU(3). 

\paragraph{The truncation scheme\\}

\begin{figure}
  \centerline{ \epsfig{file= 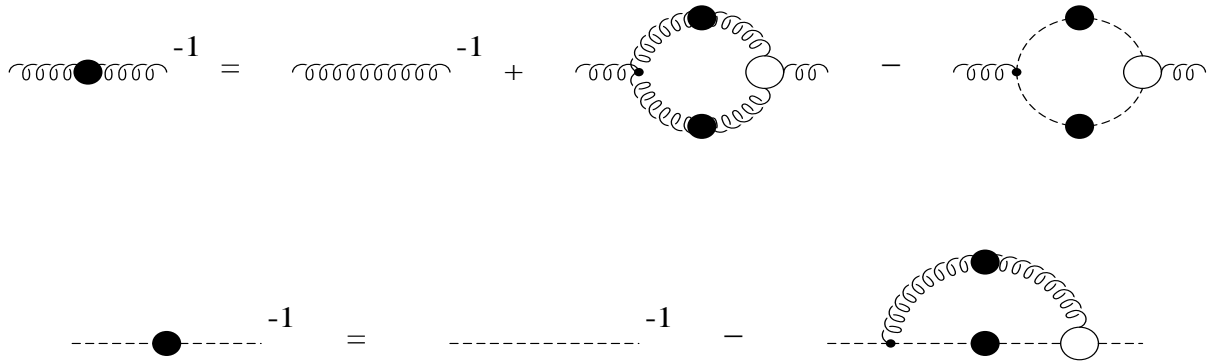,width=0.98\linewidth,height=5cm} }
  \vskip 3mm
  \caption{Diagrammatic representation of the truncated gluon and ghost 
  Dyson--Schwinger equations studied in this letter. Terms with four--gluon vertices
  have been dismissed.}
  \label{GluonGhost}
\end{figure}
To make this letter self-contained we will give a short summary of the
truncation scheme introduced in detail in ref.~\cite{Fischer:2002}. It has
turned out in previous studies \cite{Lerche:2001}, that the dressing of the
three-point vertex functions does not alter the qualitative behaviour of  the
ghost and gluon system. We therefore restrict ourselves to bare vertices for
simplicity. Furthermore all diagrams including four-gluon vertices have been
dropped. A diagrammatical  representation of the resulting system of equations
is presented in Fig.~\ref{GluonGhost}. To extract the gluon dressing function
the gluon equation has to be contracted with a tensor. Maintaining Lorentz
invariance its general form reads
\begin{equation}
{\mathcal P}^{(\zeta )}_{\mu\nu} (k) = \delta_{\mu\nu} - 
\zeta \frac{k_\mu k_\nu}{k^2} 
\, . 
\label{Paproj}
\end{equation} 
The parameter $\zeta$ allows to continuously interpolate from the transversal projector, 
$\zeta=1$, to the Brown-Pennington projector \cite{Brown:1988bm}, $\zeta=4$. In a full 
treatment without truncation the gluon equation would be transverse and the propagators 
would not depend on $\zeta$. We will see later on to what extent  transversality is violated
in the present truncation scheme.

Closely connected to the subject of transversality is the appearance of
unphysical qua\-dratic divergencies in the gluon equation. For $\zeta=4$ the
equation is projected on the  longitudinal tensor $k_\mu k_\nu/k^2$ and no
quadratic ultraviolet divergences appear \cite{Brown:1988bm}. For a transverse 
polarization the gluon equation would be independent of $\zeta$, and thus no
quadratic  divergences would occur  for all $\zeta$ in a full equation. In the
practical case, the loss of transversality causes quadratic divergences to
appear and their careful removal is essential for a numerical solution of the
system. However, this procedure should not change the infrared properties of
the equations. Anticipating that  the ghost loop is the leading contribution to
the infrared we subtract the quadratically  ultraviolet divergent constant in
the gluon loop.

The resulting system of equations for the ghost dressing function $G(x)$ and the gluon 
dressing function $Z(x)$ reads
\begin{eqnarray} 
\frac{1}{G(x)} &=& \tilde{Z}_3 - g^2N_c \int \frac{d^4q}{(2 \pi)^4}
\frac{K(x,y,z)}{xy}
G(y) Z(z) \; , \label{ghostbare} \\ 
\frac{1}{Z(x)} &=& Z_3 + g^2\frac{N_c}{3} 
\int \frac{d^4q}{(2 \pi)^4} \frac{M(x,y,z)}{xy} G(y) G(z) + Z_1
 g^2 \frac{N_c}{3} \int \frac{d^4q}{(2 \pi)^4} 
\frac{Q(x,y,z)}{xy} Z(y) Z(z) \; .\nonumber\\
\label{gluonbare} 
\end{eqnarray} 
The symbols $x:=k^2$ and $y:=q^2$ denote squared momenta
where $z=x+y-2\sqrt{xy}\cos\Theta$.
The kernels ordered with respect to powers of $z$ have the form:
\begin{eqnarray}
K(x,y,z) &=& \frac{1}{z^2}\left(-\frac{(x-y)^2}{4}\right) + 
\frac{1}{z}\left(\frac{x+y}{2}\right)-\frac{1}{4}\; , \\
M(x,y,z) &=& \frac{1}{z} \left( \frac{\zeta-2}{4}x + 
\frac{y}{2} - \frac{\zeta}{4}\frac{y^2}{x}\right)
+\frac{1}{2} + \frac{\zeta}{2}\frac{y}{x} - \frac{\zeta}{4}\frac{z}{x} \; , \\
Q(x,y,z) &=& \frac{1}{z^2} 
\left( \frac{1}{8}\frac{x^3}{y} + x^2 -\frac{19-\zeta}{8}xy + 
\frac{5-\zeta}{4}y^2
+\frac{\zeta}{8}\frac{y^3}{x} \right)\nonumber\\
&& +\frac{1}{z} \left( \frac{x^2}{y} - \frac{15+\zeta}{4}x-
\frac{17-\zeta}{4}y+\zeta\frac{y^2}{x}\right)\nonumber\\
&& - \left( \frac{19-\zeta}{8}\frac{x}{y}+\frac{17-\zeta}{4}+
\frac{9\zeta}{4}\frac{y}{x} \right) \nonumber\\
&& + z\left(\frac{\zeta}{x}+\frac{5-\zeta}{4y}\right) + z^2\frac{\zeta}{8xy} 
+ \frac{5}{4}(4-\zeta).
\end{eqnarray}
Here the last term in the gluon kernel $Q$ is introduced by hand to subtract all
quadratic divergencies from the ghost and gluon loop, see ref.~\cite{Fischer:2002}. 

In order to ensure the correct one loop scaling of the dressing functions 
in the ultraviolet within this truncation scheme
one has to violate
the Slavnov--Taylor identity $Z_1=Z_3/\tilde{Z}_3$, similar to the treatment in \cite{vonSmekal:1997is}.
Refering the reader to the detailed derivation in \cite{Fischer:2002}, we only give
the resulting momentum dependent substitution ${\cal Z}_1$ for $Z_1$:
\begin{equation} 
Z_1 \to {\cal Z}_1(x,y,z)  = 
\frac{G(y)^{(1-a/\delta-2a)}}{Z(y)^{(1+a)}}
\frac{G(z)^{(1-b/\delta-2b)}}{Z(z)^{(1+b)}}
\end{equation}
With this substitution the logarithmic behaviour of the gluon loop in the ultraviolet
is the same as the one of the ghost loop. The coefficients in the kernels then sum up to the correct 
anomalous dimensions, namely
$\delta=-9/44$ for the ghost and $\gamma=-13/22$ for the gluon propagator.
The parameters $a$ and $b$ can be chosen freely. In our calculations we use $a=b=3\delta$, which has
been shown \cite{Fischer:2002} to minimize the momentum dependence of 
${\cal Z}_1$. 
Note that $a=b=0$ leads to the truncation scheme of ref.~\cite{Bloch:2001wz} 
and $a=3\delta$, $b=0$ 
together with appropriate vertex dressings to the one of 
ref.~\cite{vonSmekal:1997is}.

\paragraph{Infrared analysis\\}

In the truncation schemes of refs.\ 
\cite{vonSmekal:1997is,Atkinson:1998tu,Atkinson:1998cz,
Zwanziger:2001kw,Lerche:2001,Fischer:2002}
the ghost loop provides the infrared leading term in the gluon equation. The
only exception is given in ref.\ \cite{Bloch:2001wz}: in this scheme also the
gluon loop contributes to the same order as the ghost loop. As we do not want to
change the infrared behaviour by manipulations of the renormalization we choose
$a=b=3\delta$ which is the only choice to give a constant ${\cal Z}_1$ in the 
infrared. Then the gluon loop does not
contribute to the infrared behaviour of the equations. Therefore the leading
analytical solutions found in the ghost loop only truncations of
\cite{Atkinson:1998cz} with $\zeta=4$ and of
\cite{Zwanziger:2001kw,Lerche:2001} with $\zeta=1$, are found in our case as
well \cite{Fischer:2002}.   In the following we will determine subleading
contributions in the infrared, and judge their importance for the numerical
treatment of the coupled system of equations.

We employ the ansatz
\begin{eqnarray}
Z(x) &=& A x^{2 \kappa} (1+f x^{\rho}) \nonumber\\
G(x) &=& B x^{- \kappa} (1+g x^{\rho}) 
\label{IRansatz}
\end{eqnarray}
in the eqs.~(\ref{ghostbare}) and (\ref{gluonbare}). The left hand side and the integrands
are expanded to appropriate powers of momentum. After integration 
(for technical details see
refs.~\cite{Lerche:2001,Fischer:2002,Hauck:1998sm}) 
the coefficients of equal powers on both sides of the 
equations have to match for consistency. The conditions on the leading term 
remain unchanged. Hence for an arbitary projector the leading power $\kappa$
is given by
\begin{equation}
\frac{(2+\kappa)(1+\kappa)}{18(3-2\kappa)}
-\frac{4\kappa-2}{4\zeta\kappa-4\kappa+6-3\zeta}=0 \; .
\label{kappa}
\end{equation}
This equation has at least two solutions for a given $\zeta$ from which only one connects
to the numerical solution for finite momenta \cite{Fischer:2002}. The result for the running 
coupling is given by
\begin{equation}
\alpha(0)=\frac{4 \pi}{18}\frac{\Gamma(3-2\kappa)\Gamma(3+\kappa)\Gamma(1+\kappa)}{\Gamma^2(2-\kappa)
\Gamma(2\kappa)} \approx 2.972,
\end{equation}
where the nonperturbative definition $\alpha(x)=\alpha(\mu^2)Z(x)G^2(x)$
has been used.
  
Matching subleading powers leads to the coupled set of homogeneous equations for $f,g$ and $\rho$:
\begin{eqnarray}
%\left(\frac{3 N_c g^2 A B^2}{48 \pi^2}
\left(3 \nu 
\frac{6 \kappa (\kappa-1) (-3+2\kappa)}{(\kappa+\rho-2)(\kappa+\rho-1)(\kappa+\rho)}
\frac{\Gamma(2\kappa)\Gamma(2-\kappa+\rho)\Gamma(3-\kappa-\rho)}
{\Gamma(4-2\kappa)\Gamma(1+\kappa-\rho)\Gamma(3+\kappa+\rho)}
-1\right) g && \hspace{1.3cm}\nonumber\\
% + \frac{3 N_c g^2 A B^2}{48 \pi^2}
+ \left( 3 \nu
\frac{3 (-2+2\kappa+\rho)}{2(\kappa+\rho-2)(\kappa+\rho-1)(\kappa+\rho)}
\frac{\Gamma(2-\kappa)\Gamma(2\kappa+\rho+1)\Gamma(3-\kappa-\rho)}
{\Gamma(3-2\kappa-\rho)\Gamma(1+\kappa)\Gamma(3+\kappa+\rho)}
\right) f &=& 0 \; , \nonumber\\
\left(\nu \frac{4 \zeta \kappa - 4\kappa + 2\rho -2 \zeta \rho +6 - 3\alpha}{2\kappa-\rho-1}
\frac{\Gamma(2-\kappa)\Gamma(2\kappa-\rho) \Gamma(2-\kappa+\rho)}
{\Gamma(1+\kappa)\Gamma(1+\kappa-\rho)\Gamma(4-2\kappa+\rho)}
\right) g + f &=& 0 \; .
\end{eqnarray}
Here $\nu={N_c g^2 A B^2}/{48 \pi^2}={\alpha(0)}/{4\pi}$. There is either the trivial solution $f=g=0$ 
or one has to set the determinant of these linear equations
to zero. We then obtain the results $\rho_{(1)}=0, \rho_{(2)}=0.58377, 
\rho_{(3)}=1.20300$ and several other solutions with higher values of the power.
The solution $\rho_{(1)}=0$ corresponds to the pure power solution. The lowest nonvanishing 
solution, $\rho_{(2)}=0.58377$, is sufficently high that we safely may neglect it in the numerical 
treatment of the infrared part of the equations. This will be detailed in the next section.

\paragraph{Numerical Method and Results\\}

%%%%%%%%%%%%%%%%%%%%%%%%%%%%%%%%%%%%%%%%%%%%%%%%%%%%%%%%%%%%%%%%%%%%%%%%%%%%%%%%%%%%%%%%%%%%%%%%%%%%%%%%%%%%%%
\begin{figure}[t]
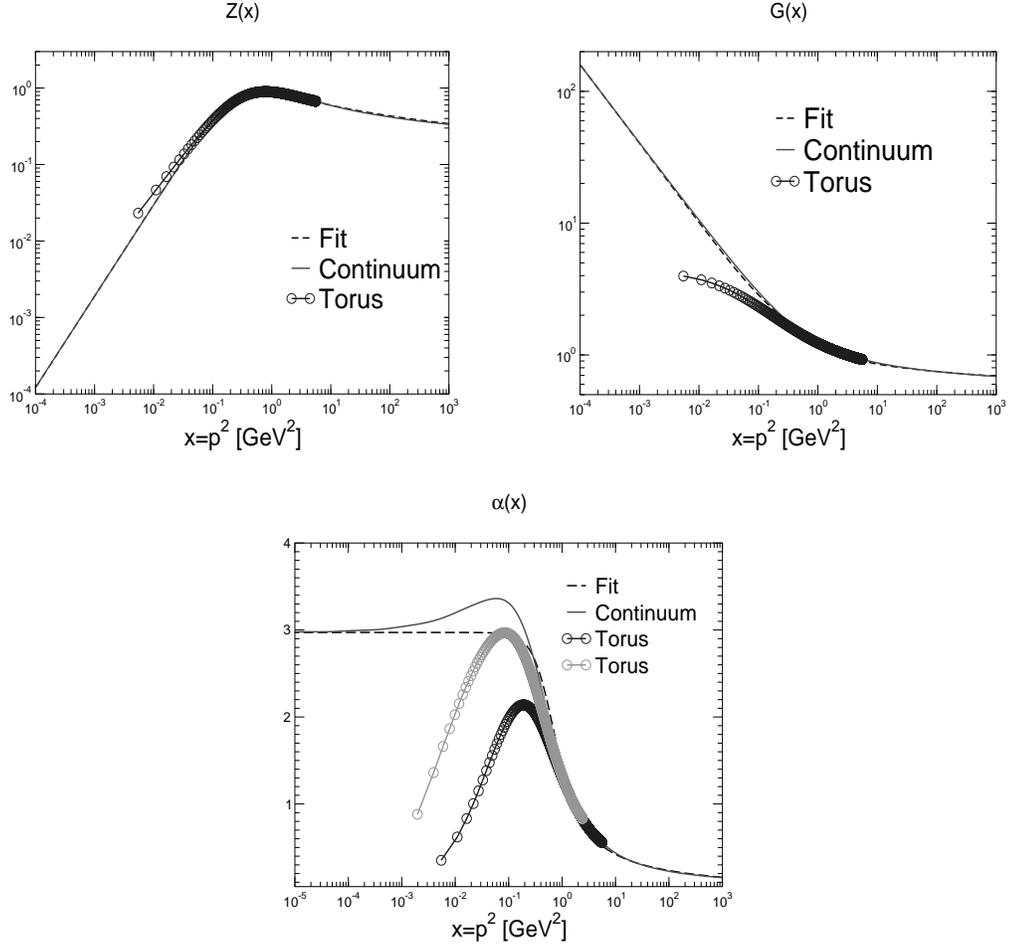

\vspace{-0cm}
\centerline{
\epsfig{file=cont.main.z.eps,width=6.0cm,height=6cm}
\hspace{1cm}
\epsfig{file=cont.main.g.eps,width=6.0cm,height=6cm}
}
\vspace{0.5cm}
\centerline{
\epsfig{file=cont.main.a.eps,width=6.0cm,height=6cm}
}
\caption{\label{new.dat}
Shown are the results for the gluon dressing function, the ghost dressing 
function and the running coupling using a transverse projector. 
The solutions are compared to previous results on a torus, for the running
coupling two results for two different volumes are presented. The dashed lines
are the fits given in eq.~(\protect{\ref{fit}}).}
\end{figure}
%%%%%%%%%%%%%%%%%%%%%%%%%%%%%%%%%%%%%%%%%%%%%%%%%%%%%%%%%%%%%%%%%%%%%%%%%%%%%%%%%%%%%%%%%%%%%%%%%%%%%%%%%%%%%%

Two of the four integrals in equations (\ref{ghostbare}) and (\ref{gluonbare})
can be done trivially and yield a factor of $4\pi$, while the remaining angular
integral and  the radial integral have to be performed with the help of
numerical routines. To achieve high accuracy we split the radial loop integral
into three parts, $y \in [0,\epsilon^2]$, $y \in (\epsilon^2,x]$ and  $y \in
(x,x_{UV}]$. The integrals will be computed numerically, however according to
the value of their argument the dressing functions $Z$ and $G$ have to be
handled differently. In the infrared region, $y,z \in [0,\epsilon^2]$, $Z$ and
$G$ behave like powers and are replaced according to equations (\ref{IRansatz})
with $f=g=0$. The infrared matching point is chosen to be $\epsilon^2=(0.55
\mbox{MeV})^2$ in our calculations. In the high momentum regime, $y \in
(x,x_{UV}]$, arguments $z$ occur which are larger than the numerical cutoff
$x_{UV}$. There we approximate the respective dressing functions by the
expressions
\begin{eqnarray}
Z(z) &=& Z(s) \left[ \omega \log\left(\frac{z}{s}\right)+1 \right]^\gamma  \; ,
\label{gluon_uv}\\
G(z) &=& G(s) \left[ \omega \log\left(\frac{z}{s}\right)+1 \right]^\delta  \; .
\label{ghost_uv}
\end{eqnarray}
according to the one loop behaviour of the solutions as has been detailed in \cite{Fischer:2002}. Here
$\omega = 11N_c\alpha(s)/12\pi$ and the squared momentum $s$ is a perturbative scale. We chose 
$s=(174 \mbox{GeV})^2$ to be slightly lower than the numerical cutoff $x_{UV}=(177 \mbox{GeV})^2$.   
To be able to perform the angular integrations for momenta $[\epsilon^2,x_{UV}]$
we expand the dressing functions in Chebychev 
polynomials and solve the coupled system of equations for the expansion
coefficients using a Newton iteration
method. Details of this technique can be found in ref.\ \cite{Atkinson:1998tu}. 

We apply a MOM regularization scheme similar to the ones used previously in 
refs.~\cite{vonSmekal:1997is,Atkinson:1998tu}. For numerical reasons it is favourable
to subtract  the ghost equation at zero momentum and the gluon equation at the
scale $s$ introduced above. Due to this the unknown renormalization constants
$Z_3$ and $\tilde{Z}_3$ drop out and instead of them one has to specify two
input variables. We chose the infrared parameter $A$ 
({\it c.f.}~eq.~(\ref{IRansatz})), and implement the condition 
$Z(\mu^2)G^2(\mu^2)=1$. 
Furthermore, one has to  specify the coupling $\alpha(\mu^2)=g^2/16\pi^2$
entering eqs.\ (\ref{ghostbare},\ref{gluonbare}). 
Finally we have fixed the momentum scale by calculating
the running coupling for the colour group SU(3), and requiring 
the experimental value
$\alpha(x)=0.118$ at $x=M_Z^2=(91.187 \mbox{GeV})^2$ \cite{pdg}. 

%%%%%%%%%%%%%%%%%%%%%%%%%%%%%%%%%%%%%%%%%%%%%%%%%%%%%%%%%%%%%%%%%%%%%%%%%%%%%%%%%%%%%%%%%%%%%%%%%%%%%%%%%%%%%%
\begin{figure}[t]
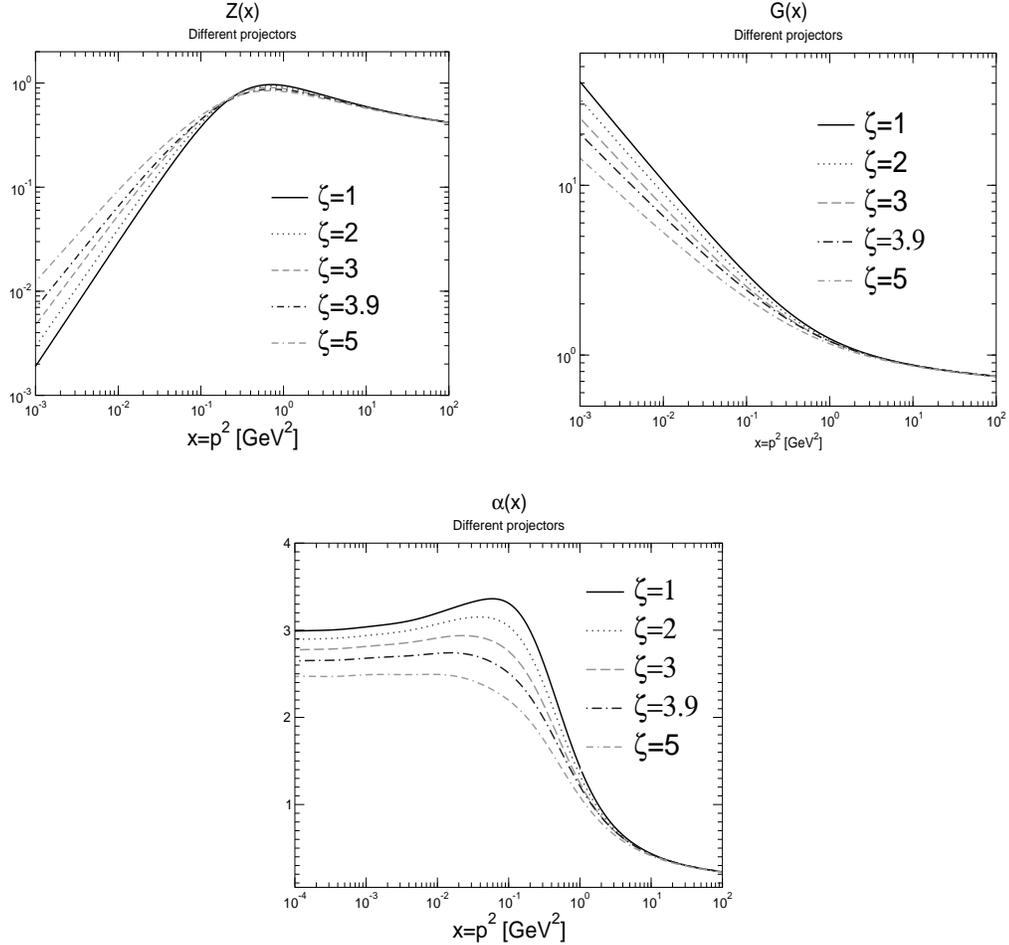

%\vspace{-1cm}
\centerline{
\epsfig{file=cont.proj.z.eps,width=6.0cm,height=6cm}
\hspace{1cm}
\epsfig{file=cont.proj.g.eps,width=6.0cm,height=6cm}
}

\vspace{0.5cm}
\centerline{
\epsfig{file=cont.proj.a.eps,width=6.0cm,height=6cm}
}
\caption{\label{proj.dat}
Shown are the results for the gluon dressing function, the ghost dressing 
function and the running coupling, {\it c.f.}~fig.\ \protect{\ref{new.dat}},
for different projectors.}
\end{figure}
%%%%%%%%%%%%%%%%%%%%%%%%%%%%%%%%%%%%%%%%%%%%%%%%%%%%%%%%%%%%%%%%%%%%%%%%%%%%%%%%%%%%%%%%%%%%%%%%%%%%%%%%%%%%%%

In Fig.~\ref{new.dat} we show the results for the transverse projector,
$\zeta=1$. The gluon and ghost dressing functions behave powerlike for low
momenta and obey one loop scaling   in the ultraviolet as expected. 
Accordingly the running coupling has a fixed point in the infared and decreases
logarithmically in the perturbative regime above several GeV. For intermediate
momenta the behaviour of the running coupling seems to suggest a second zero of
the $\beta$-function around  $(100 \mbox{MeV})^2$. However,  such an extremum
in $\alpha (x)$ would result in a double valued $\beta$-function, therefore  we regard it
as an artefact of our truncation scheme. In fact with subleading gluon
contributions in the infrared and perturbative suppression of the two loop
contributions in the ultraviolet it is exactly the intermediate momentum regime
where the omission of the two loop diagrams should cause deficiencies in our
solutions.

The asymptotic behaviour of the solutions can also be seen from the
functional form of our fits. The dressing functions $Z(x)$ and $G(x)$ can be described by
\begin{eqnarray}
\alpha(x) = \frac{\alpha(0)}{\ln(e+a_1 x^{a_2}+b_1x^{b_2})}, \:\:\: && 
R(x) = \frac{c x^{\kappa}+dx^{2\kappa}}{1+ c x^{\kappa}+dx^{2\kappa}}, \nonumber\\
Z(x) = \left( \frac{\alpha(x)}{\alpha(\mu)} \right)^{1+2\delta} R^2(x), \:\:\: &&
G(x) = \left( \frac{\alpha(x)}{\alpha(\mu)} \right)^{-\delta} R^{-1}(x), 
\label{fit}
\end{eqnarray}
with fitting parameters $a_1,a_2,b_1,b_2$ and $c,d$ for the running coupling
$\alpha(x)$ and the auxiliary function $R(x)$ respectively. The values of
$\alpha(0)=2.972$ and $\kappa=0.5953$ are known from the infrared analysis in
the previous paragraph, whereas $\delta=-9/44$ is the anomalous dimension of
the ghost dressing function and $\alpha(\mu^2)=0.9676$.
The six parameters of the fit are given by 
$a_1=5.292 \mbox{GeV}^{-2a_2}$, $a_2=2.324$,  
$b_1=0.034 \mbox{GeV}^{-2b_2}$, $b_2=3.169$,
$c=1.8934 \mbox{GeV}^{-2\kappa}$ and  $d=4.6944 \mbox{GeV}^{-4\kappa}$. As can
be seen in Fig.~\ref{new.dat} the fit works very well and can be used as input
for phenomenological calculations in future work.     

Our results for different
values of the parameter $\zeta$ ({\it c.f.}~eq.~(\ref{Paproj})) are shown in Fig.~\ref{proj.dat}. In accordance
with the infrared analysis the power $\kappa$ changes from $\kappa=0.5953$ for $\zeta=1$ to $\kappa=0.4610$
for $\zeta=5$. The perturbative properties of the solutions remain unchanged. The bump in the
running coupling gets smaller but does not disappear even for $\zeta=5$. It has already been stated above
that the dressing functions would be independent of $\zeta$ in a complete treatment of the gluon equation.
As all our solutions are very similar even on a quantitative level we conclude that transversality is 
lost only to a moderate extent. This is a somewhat surprising result in such a simple truncation scheme 
as the one at hand.

The Brown--Pennington projector, $\zeta=4$, is an exceptional case as can be
seen from eq.~(\ref{kappa}).  Here the $\kappa$-dependence of the second term
cancels and only one solution, $\kappa=1$, can be found ({\it
c.f.}~ref.~\cite{Atkinson:1998cz}). We found no numerical solutions for this
case. However, within the limit of numerical accuracy, solutions for $\zeta$
slightly different from 4 can be found leading to a value for $\kappa$ slightly
different from 1/2. {\it E.g.} in Fig.~\ref{proj.dat} the case $\zeta=3.9$
leading to $\kappa=0.5038$  is depicted.

%%%%%%%%%%%%%%%%%%%%%%%%%%%%%%%%%%%%%%%%%%%%%%%%%%%%%%%%%%%%%%%%%%%%%%%%%%%%%%%%%%%%%%%%%%%%%%%%%%%%%%%%%%%%%%
\begin{figure}[t]
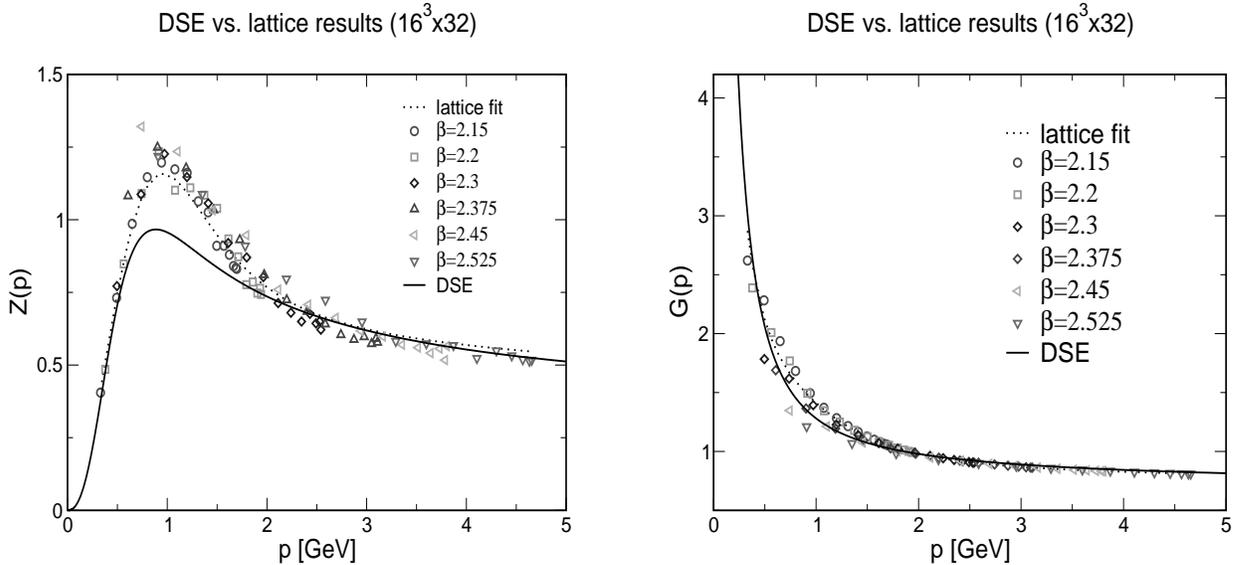

%\vspace{-2cm}
\centerline{
\epsfig{file=lattice_gluecont.eps,width=7.5cm,height=7.5cm}
\hspace{1.0cm}
\epsfig{file=lattice_ghostcont.eps,width=7.5cm,height=7.5cm}
}
\caption{\label{lattice.dat}
Solutions of the Dyson-Schwinger equations compared to recent lattice results 
for two colours \cite{Langfeld:2002}. 
%The differences at intermediate momenta indicate deficiencies of the truncation scheme.
}
\end{figure}
%%%%%%%%%%%%%%%%%%%%%%%%%%%%%%%%%%%%%%%%%%%%%%%%%%%%%%%%%%%%%%%%%%%%%%%%%%%%%%%%%%%%%%%%%%%%%%%%%%%%%%%%%%%%%%

Finally, we compare our results to recent lattice calculations for two colours 
\cite{Langfeld:2002}. As the solutions on the lattice include all
non-perturbative effects,  the results shown in Fig.~\ref{lattice.dat} confirm
again that the omission of the two loop diagrams in our truncation scheme
mostly effects the region around the bending point at $1$ GeV. However, in 
general the qualitative and partly even quantitative agreement of the two
methods is remarkable. The combined evidence of the two methods points
strongly towards an infrared vanishing gluon propagator and an infrared
singular ghost propagator in Landau gauge. 

\paragraph{Conclusions\\}

In this letter we have presented approximate non-perturbative solutions for the
gluon and the ghost  propagators as well as the running coupling in Landau
gauge. We obtained these solutions for the Dyson--Schwinger equations in the
truncation of ref.~\cite{Fischer:2002}, working with bare vertices and omitting
all diagrams that involve four-gluon vertices. An improvement to previous
treatments has been the explicit numerical  calculation  of all angular
integrals thus overcoming the angular approximations that have been made so
far. It is interesting to note that only one solution (for a given projector)
has been found. 
Furthermore, we could demonstrate that our truncation scheme violates
transversality of the  gluon equation only to a moderate extent. Previous
findings on a four-torus have been checked and found to be in agreement with
the flat space-time solutions up to effects of finite volume in the very
infrared. Despite the simplicity of the truncation our solutions agree
remarkably well with recent lattice calculations. For the infrared behaviour of
Landau gauge Green's functions  circumstancial evidence points towards a weakly
infrared vanishing or even finite gluon propagator, while the ghost propagator
seems to rise more strongly than a simple pole.   

\paragraph{Acknowledgements\\}

We thank Jacques Bloch, Hugo Reinhardt, Sebastian Schmidt,
Peter Watson and Daniel Zwanziger for helpful discussions.
We are especially grateful to Lorenz von Smekal for many valuable discussions and
for making ref.\ \cite{Lerche:2001} available to us. We are indebted to Kurt Langfeld
for communicating and elucidating his lattice results, partly prior to publication.

This work has been supported by the DFG under contract Al 279/3-3 and by the
European graduate school T\"ubingen--Basel.

\end{document}